\documentclass[aip,graphicx]{revtex4-1}
\usepackage{amssymb}

\usepackage{graphicx}

\draft

\begin{document}

\title{Designing phase-sensitive tests for Fe-based superconductors}
\author{A.A.~Golubov}
\affiliation{Faculty of Science and Technology and MESA+ Institute of
Nanotechnology, University of Twente, 7500 AE Enschede, The Netherlands}

\author{I.I.~Mazin}
\affiliation{Naval Research Laboratory, 4555 Overlook Ave. SW, Washington,
DC 20375, USA}

\date{\today}

\begin{abstract}
We suggest new experimental designs suitable to test pairing symmetry in
multiband Fe-based superconductors. These designs are based on combinations
of tunnel junctions and point contacts and should be accessible by existing
sample fabrication techniques.
\end{abstract}

\maketitle





Four years after the discovery of the new family of high-$T_{c}$ Fe-based
superconductors (FeBS) \cite{kamihara}, their pairing symmetry is still
under dispute\cite{ROPP}. While most researchers favor the so-called $s_{\pm
}$ pairing, whereupon the sign of the order parameters changes between the
hole and the electron bands \cite{mazin}, some advocate \cite{Kontani} the
more conventional anisotropic $s$, and for the extreme cases such as KFe$_{2}
$As$_{2}$ and K$_{x}$Fe$_{2}$Se$_{2}$ other alternatives have been suggested
(d-wave, or other types of sign-changing $s$). This reminds us of the
controversy in the high-T$_{c}$ cuprates, when proponents and deniers of the
d-wave pairing were clinched in dead heat for several years, until the first
phase-sensitive tunneling experiments had been performed\cite%
{wollman,tsuei,harlingen,kirtley}, and showed unambigously that the
Josephson current flowing from a cuprate sample along the $y$ direction is
shifted by $\phi =\pi $ with respect to the corresponding current flowing in
the $x$ direction.

Despite recent progress in junction fabrication \cite{zhang,zhou}, no such
(or similar) phase-sensitive experiments have been performed so far in
FeBS-based Josephson junctions, designed and produced in a controllable way.
Only indirect evidence that Josephson loops with a $\pi $ phase shift can be
formed in these materials was reported in Ref. \cite{chen} where samples
with a large number of randomly formed contact pairs were measured.

Apart from problems with sample preparations, and other technical obstacles,
a serious barrier preventing similar decisive experiments in FeBS is the
fact that the two main contenders for the pairing state in the
\textquotedblleft mainstream\textquotedblright\ FeBS are $s_{\pm }$ and $%
s_{++},$ two states that have the same orbital symmetry. Therefore one needs
to design the experimental geometry in a particularly clever way so that the
current in one contact would be dominated by the carriers having one sign of
the order parameter, and in the other by carriers with the opposite sign.
Note that designing the Josephson contacts so that current would be flowing
in different Cartesian directions is not necessary, and in fact not helpful
at all, because an $s$-wave superconductor is invariant under the x-y
rotation.

Several designs aimed at exploiting particular Fermi surface topology of
FeBS have been suggested, such as placing contacts at an angle different
from 90$^{\circ },$ or below and above a sandwich of two different
superconductors \cite{parker,wu}. All these suggestions have proven to be
too complicated to be realized in practice. In this Letter we suggest three
new experimental designs, all of them much simpler than all proposed
previously. All these designs should be accessible by available experimental
techniques and existing sample manufacturing is already at a level
sufficient for exploiting these new ideas.

Before describing our suggestions in details, we would like to make a
general observation that in fact allowed us to come up with the designs so
much simpler than those discussed previously. There is a powerful tool in
our hands, namely, a choice between planar tunnel junctions, where the
current is dominated by the electrons with the momentum normal to the
interface, and point contacts that collect the current indiscriminately from
all electrons.

Let us elaborate more on the first point.

For planar tunnel junctions with a thick specular barrier electrons
tunneling normal to the interface have an exponentially big advantage over
those with a finite momentum parallel to the interface, $k_{\parallel }\neq
0.$ For instance, the tunneling probability $T_{\mathbf{k}}$ for a simple
vacuum barrier can be expressed as\cite{mazin_epl_prl}
\begin{equation}
T_{\mathbf{k}}=\frac{4m_{0}^{2}\hbar ^{2}K^{2}v_{L}v_{R}}{\hbar
^{2}m_{0}^{2}K^{2}(v_{L}+v_{R})^{2}+(\hbar
^{2}K^{2}+m_{0}^{2}v_{L}^{2})(\hbar ^{2}K^{2}+m_{0}^{2}v_{R}^{2})\sinh
^{2}(dK)}.
\end{equation}%
Here $m_{0}$ is the electron mass, $v_{L,R}$ are the Fermi velocity
projections on the tunneling directions, $d$ is the width of the barrier,
and the quasimomentum of the evanescent wavefunction in the barrier, $iK,$
is, from the energy conservation,%
\begin{equation}
K=\sqrt{k_{\parallel }^{2}+2(U-E)m_{0}},
\end{equation}%
where $U$ is the barrier height.

The Josephson current in such tunnel junction between a single- and
multi-band superconductor is determined by a standard Ambegaokar-Baratoff
formula
\begin{equation}
I_{S}=\frac{\pi T}{eR_{0}}\sum\limits_{n,i=1,2}\frac{\Delta _{L}\Delta
_{R}\sin \phi }{\sqrt{\omega _{n}^{2}+\Delta _{L}^{2}}\sqrt{\omega
_{n}^{2}+\Delta _{R}^{2}}},
\end{equation}%
where $\Delta _{L}$ is the gap in a single-band superconductor, $\Delta _{R}$
is the gap in a multi-band superconductor corresponding to a Fermi surface
sheet in the center of the Brillouin zone and $R_{0}$ is the corresponding
tunneling resistance, controlled by small values of $k_{\parallel }.$

In the opposite regime of a point contact (ScS-type) between a single band
superconductor and a multiband one, there is no conservation of $%
k_{\parallel }$ (in fact, it is not even well-defined) and essentially all
electrons contribute to the total current through the contact. This
situation can be modelled by a diffusive contact that does not respect
momentum conservation. Then the relative contribution to supercurrent from
band `$i$' is determined by the partial resistance $%
R_{Ni}^{-1}=(2Se^{2}/L)N_{i}D_{i},$ \cite{Omel} where $N_{i}$, $D_{i}$ are
densities of states and diffusion coefficients in the corresponding band, $L$
and $S$ are the length and crossection area of a contact. This amounts to
adding all conductivity channels for each direction independently, resulting
in the DOS-weighted average of the corresponding squared Fermi velocity, $%
e.g.,$ $\left\langle N(E_{F})v_{F}^{2}\right\rangle .$ In the practically
relevant case when\ $\Delta _{L}\ll \Delta _{Ri}$ the Josephson current in a
diffusive ScS contact between a single- and a two- band superconductors is
is given by the following simple expression
\begin{equation}
I_{S}=\frac{\Delta _{L}}{e}\sum\limits_{i=1,2}\left[ \ln \frac{\Delta
_{Ri}\cos \phi /2}{\Delta _{L}(1+\cos \phi )}\right] \frac{\sin \phi }{R_{Ni}%
},
\end{equation}%
which is a multiband generalization of the well known formula (see, $e.$ $g.,
$ Refs. \cite{KO,Omel,RMP}). From this formula, it follows, with
logarithmic accuracy, that current-phase relation is sinusoidal with
critical current controlled by the corresponding resistance $R_{Ni}$ only.

Based on the theoretical consideration above, we want to suggest three new
experimental designs.

\begin{figure}[tbh]
\begin{center}
\includegraphics[width=1\columnwidth]{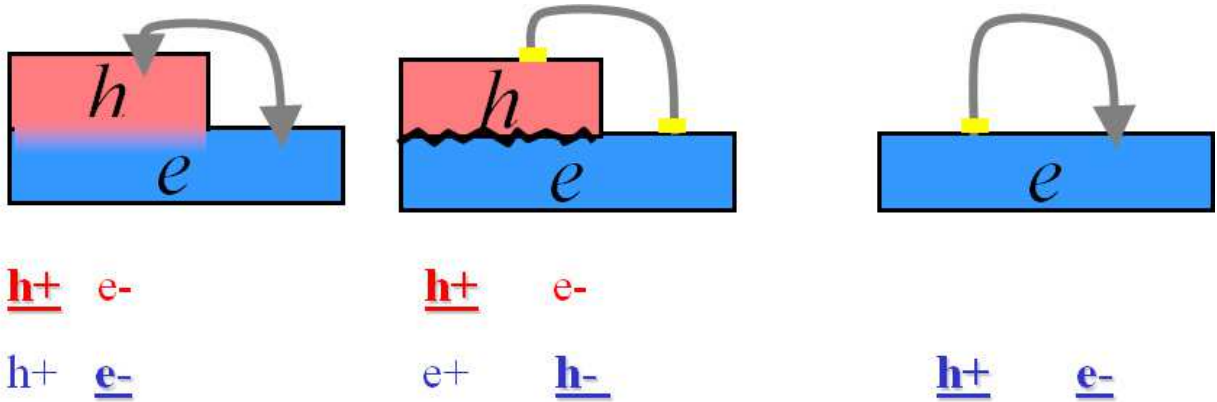}
\end{center}
\caption{Suggested experimental designs of Josephson $\protect\pi$-loops:
epitaxial sandwich (left); rough sandwich (middle); single sample (right) }
\label{design}
\end{figure}

\begin{table}[tbp]
\caption{Three suggested designs for probing the relative phases of the
order parameter in Fe-based superconductors. A tunneling barrier here is
assumed to be thick enough to filter through the \textquotedblleft tunneling
cone\textquotedblright\ effect only the states near the zone center (holes),
while a point contact is supposed to collect current in all directions and
thus be dominated by the majority carriers. The sign of the order parameter
is selected in such a way that the current through the left (upper) contact
is always considered positive. }
\label{table_stab}%
\begin{tabular}{|l|c|c|c|}
\hline
Design &  &  &  \\
Fig. 1 panel & left & middle & right \\
Upper/left contact & point & tunnel & tunnel \\
Lower/right contact & point & tunnel & point \\
Upper $\Delta _{hole}$ & $-$ & + & + \\
Upper $\Delta _{elec}$ & + & $-$ & $-$ \\
Interface & epitaxial & rough & n/a \\
Lower $\Delta _{hole}$ & $-$ & $-$ & n/a \\
Lower $\Delta _{elec}$ & + & + & n/a \\
Upper contact current dominated by & electrons & holes & holes \\
Lower contact current dominated by & holes & holes & electrons%
\end{tabular}%
\end{table}

\textit{1. Epitaxial sandwich.} Here we propose to grow an electron-doped
film (for instance, Co-doped BaFe$_{2}$As$_{2}),$ and on top of this film,
as shown in Fig. \ref{design}, to grow epitaxially a hole-doped film
(K-doped BaFe$_{2}$As$_{2}).$ Epitaxially grown films (there is hardly any
lattice mismatch between the optimally doped K$_{x}$Ba$_{1-x}$Fe$_{2}$As$_{2}
$ and optimally doped BaCo$_{x}$Fe$_{2-x}$As$_{2})$ conserves the lateral
translational symmetry, and therefore the electron momentum parallel to the
interface is also conserved. This means that the conductance between the sandwich
buns is dominated by the electron-electron and hole-hole currents, while the
electron-hole and hole-electron conversion, requiring a lateral momentum
transfer of the order of $\hbar \pi /a,$ will be suppressed.

Maximizing the Josephson energy at this epitaxial interface, we have to
assign the same phases to the electron Fermi surfaces in both films, and the
opposite phase to the hole Fermi surfaces. We now close the loop by
attaching to the two films, as shown in Fig. \ref{design}, point contacts
made out of a conventional superconductor. As discussed above, the current
through a point contact is averaged over all electrons. Moreover, FeBS being
quasi-2D metals, most of the current will be flowing in the ab plane, since
a point contact penetrates inside the the bulk of the film. Now, the current
from the electron doped film into the point contact will be dominated by the
electron Fermi surfaces, simply because these carriers dominate the bulk,
and the current from the hole-doped film will be dominated by holes. These
two currents will thus have the opposite signs, or the phase shift of $\pi .$

\textit{2. Rough sandwich.} Here we suggest to physically combine two single
crystals, or two films, without creating epitaxial contact between the two.
Now our goal is to create a rough interface where the lateral momentum is
not conserved at all, and any state in the electron-doped part of the sample
can tunnel into any state of the hole part. In fact, a rough interface can
be substituted by a thin layer of a conventional superconductor with no
lattice matching to the FeBS, if that is more feasible experimentally. But,
as long as we have created a contact between the two FeBS without momentum
conservation, the current in this contact will be controlled by the majority
carriers in each electrode, so that the holes in the hole-doped part will be
in phase coherence with the electrons in the electron-doped part (to minimize the Josephson
energy).

Now we need to attach contacts to a conventional superconductor in such a
way that the current in both will be dominated by holes, even in the part
that is electron-doped, since now holes in the two electrodes have
superconducting order parameters of the opposite signs. This can be achieved
by using a planar junction with a sufficiently thick tunneling barrier in both
contacts. As discussed above, a conventional planar tunneling barrier selects
exponentially electrons with the momentum $\hbar \mathbf{k}$ such that $%
k_{\parallel }\sim 0,$ where $k_{\parallel }\sim 0$ is the projection on the
interface plane. This condition filters out electron states near the corner
of the Brillouin zone and lets trough only the hole states. Since in this
design the phase coherence between the hole and the electron doped
electrodes is between the carriers of the opposite character, we achieve a
Josephson loop with a $\pi $ shift between the contacts.

\textit{3. Single sample.} The previous two designs relied on manufacturing
a composite sample where the two contacts will be attached to two parts with
different properties. In our last design, the job of creating a phase shift
between the contacts is relegated to the difference in contacts themselves.
Here we propose a single sample (which can be a single crystal or a thin
film), to which two contacts of different nature are attached. Importantly,
the sample must be electron-doped, so that the normal current (and, by
implication, the current through a point contact) would be dominated by
electrons. We use one point contact, and one planar thick-barrier tunnel
junction with the current direction along $z.$ As discussed above, the
former will be dominated by electrons and the latter by holes, which have
small $k_{\parallel },$ thus again creating a $\pi $ shift.

\begin{figure}[tbh]
\begin{center}
\includegraphics[width=1\columnwidth]{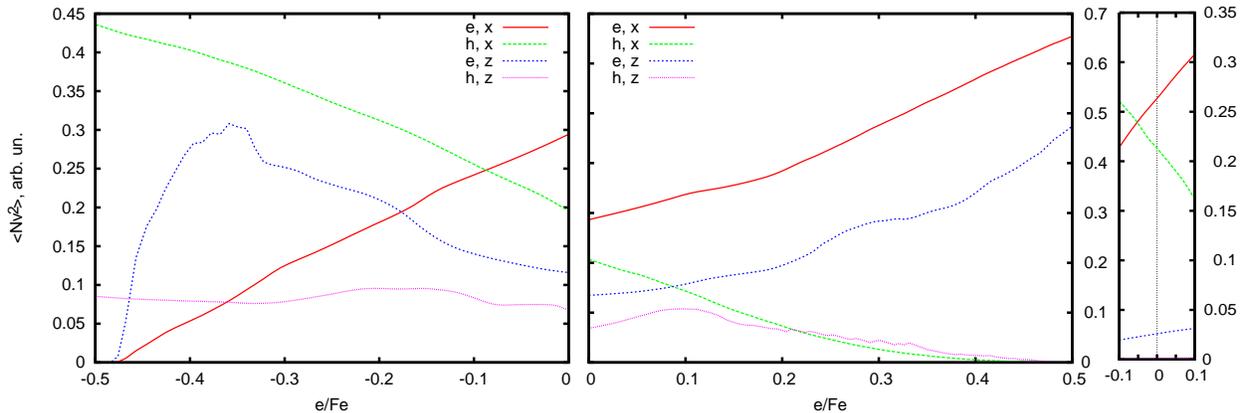}
\end{center}
\caption{Calculated transport function, $\left\langle
N(E_{F})v_{F}^{2}\right\rangle ,$ for the hole-doped (left panel) and
electron-doped (middle panel) BaFe$_2$As$_2$. Calculation for the
electron-doped case were self-consistent in the virtual crystal
approximation for the 10\% Co doping and the rigid band approximation used
around this composition. Similarly, the hole-doped composition were
self-consistent for the 40\% K doping and the rigid bands used thereafter.
The right panel shows similar calculations for FeSe. \newline
}
\label{vel}
\end{figure}

In all three designs discussed above a $\pi $ shift can be detected by
combining the contacts into a two-junction interferometer with critical
current $I_{c}=\sqrt{I_{c1}^{2}+I_{c2}^{2}\pm 2I_{c1}I_{c2}\cos 2\pi \Phi
/\Phi _{0}}$. Here $I_{c1,2}$ are critical currents of individual junctions,
$\Phi $ is magnetic flux through the interferometer, $\Phi _{0}$ is flux
quantum and sign + (-) corresponds to zero ($\pi $) shift between the
contacts.\ In such interferometer a $\pi$-shift shows up as a minimum of $I_{c}$
at $\Phi =0$ (the so-called $\pi $-SQUID behavior). In is important to note
that to observe significant $I_{c}(\Phi )$ modulation, the critical currents
$I_{c1,2}$ (and thus junctions resistances) should be of similar order of
magnitude. Tunnel junctions have much higher specific barrier resistance $%
R_{0}S$ than that in PC's, therefore in our last design (3) tunnel contact should have large enough area to
fulfill the above condition.

Finally, one may ask a question: our proposals are based on the
assumption that  the
normal (diffusive) transport in electron and hole doped FeBS is dominated by
the carriers of the corresponding sign; to what extent this assumption
is justified in actual material?
 To answer this question
we have performed the standard LAPW band structure calculations \cite{calc} and have
computed the relevant quantity\cite{note}, $\left\langle
N(E_{F})v_{F}^{2}\right\rangle ,$ as a function of doping (in the rigid band
approximation, which is enough for our qualitative purpose). The results are
shown in Fig. \ref{vel}. As one can see, the condition that the diffusive
current for electron-doped Ba122 material is dominated by electrons is well
satisfied for both in-plane ($x)$ and out-of-plane ($y)$ directions,
particularly well for overdoped ($\gtrsim 10\%)$ samples (which are
therefore preferable). The condition that for the hole doping the current be
dominated by holes is less well fulfilled. Indeed, for optimal (0.2 hole/Fe)
and even overdoped samples the current in the $z$ direction is still
dominated by electrons, because the electron Fermi surfaces are more warped.
However, the in-plane current is firmly dominated by holes for all
composition with higher than 20\% K content. Thus, the recommendation in
this case is to manufacture a point contact that penetrates into the sample
deep enough to probe the in-plane conductivity as much as the out-of-plane
one. In that case the dominance of the hole current will be assured.

Since our third design does not require a combination of two different
materials, it is interesting to check whether one can make the same
experiment with an undoped compound. Indeed, one of the most popular FeBS,
particularly in terms of thin film manufacturing, is FeSe. For comparison,
we show in the right panel of Fig. \ref{vel} the corresponding data for FeSe
with minimal doping. One can see that while for the undoped composition the
current is dominated by electrons, the effect is small and probably not
sufficient to create a good Josephson $\pi $-loop. Instead, for this
material, one should use, instead of a point contact, a thin specular
tunneling barrier (planar junction). In that case only the current in the
normal ($z)$ direction will be relevant, and, as one can see from Fig. \ref%
{vel}, this current is completely dominated by electrons\cite{note2}.

To conclude, we have suggested three new experimental designs in order to
test pairing symmetry in FeBS. These designs involve Josephson two-junction
interferometers where current in different contacts is dominated by
different type of carriers, electrons or holes. If pairing symmetry is of
the $s_{\pm }$-type, a Josephson $\pi $-loop is realized ($\pi $-SQUID),
while in the $s_{++}$ case the standard SQUID behavior is expected. The
suggested designs should be accessible by available fabrication techniques
and should allow to probe pairing symmetry in FeBS.

We thank A. Brinkman, G. Pepe and Y. Tanaka for useful discussions and acknowledge
financial support from Dutch FOM and EU-Japan program 'IRON SEA'.


\begin{thebibliography}{99}
\bibitem{kamihara} Y. Kamihara, T. Watanabe, M. Hirano, and H. Hosono, J. Am. Chem. Soc. \textbf{130}, 3296
(2008).

\bibitem{ROPP} P.J. Hirschfeld, M.M. Korshunov, I.I. Mazin, Reports on
Progress in Physics \textbf{74}, 124508 (2011).

\bibitem{mazin} I.I. Mazin, D.J. Singh, M.D. Johannes, and M.-H. Du, Phys. Rev. Lett. \textbf{101},
057003 (2008).

\bibitem{Kontani} H. Kontani and S. Onari, Phys. Rev. Lett. \textbf{104},
157001 (2010)

\bibitem{wollman} D.A. Wollman, D. J. Van Harlingen, W. C. Lee, D. M. Ginsberg, and A. J. Leggett, Phys. Rev. Lett. \textbf{71},
2134 (1994).

\bibitem{tsuei} C.C. Tsuei, J. R. Kirtley, C. C. Chi, Lock See Yu-Jahnes, A. Gupta, T. Shaw, J. Z. Sun, and M. B. Ketchen, Phys. Rev. Lett. \textbf{73}, 593 (1994).

\bibitem{harlingen} D.J. Van Harlingen, Rev. Mod. Phys. \textbf{67}, 515
(1995).

\bibitem{kirtley} C. C. Tsuei and J. R. Kirtley, Rev. Mod. Phys.\textbf{72}, 969
(2000).

\bibitem{zhou} Y.-R. Zhou, Yan-Rong Li, Jun-Wei Zuo, Rui-Yuan Liu, Shao-Kui Su, G. F. Chen, J. L. Lu, N. L. Wang, and Yun-Ping Wang, arXiv:0812.3295.

\bibitem{zhang} X. Zhang, Yoon Seok Oh, Yong Liu, Liqin Yan, Kee Hoon Kim, Richard L. Greene, Ichiro Takeuchi, arXiv:0812.3605.

\bibitem{chen} C.-T. Chen, C.C. Tsuei, M.B. Ketchen, Z.-A. Ren, and Z.X.
Zhao, Nature Physics \textbf{6}, 260 (2010).

\bibitem{parker} D. Parker and I.I. Mazin, Phys. Rev. Lett. \textbf{102}, 227007
(2009).

\bibitem{wu} J. Wu and P. Phillips, Phys. Rev. B \textbf{79}, 092502 (2009).

\bibitem{mazin_epl_prl} I.I. Mazin, Phys. Rev. Lett. \textbf{83}, 1427 (1999); Europhys. Lett. \textbf{55}, 404 (2001).

\bibitem{Omel} A. Brinkman, A.A. Golubov, and M.Yu. Kupriyanov, Phys. Rev. B \textbf{69}, 214407 (2004);
 Y.S. Yerin and A.N. Omelyanchuk, Low Temp. Phys. \textbf{36}, 969 (2010).

\bibitem{KO} I.O.Kulik and A.N. Omelyanchuk, Pis'ma Zh. Eksp.Teor. Fiz.
\textbf{21}, 216 [JETP Lett. \textbf{21}, 96 (1975)].

\bibitem{RMP} A.A. Golubov, M.Yu. Kupriyanov, and E. Ilichev, Rev. Mod.
Phys. \textbf{76}, 411 (2004).


\bibitem{calc} We used the linear augmented plane wave method (LAPW) in the
virtual crystal approximation, as discussed in Ref. \cite{mazin}. So far
experimental evidence has agreed favorably with DFT calculations. It is
generally believed that up to a moderate renormalization of the bandwidth,
DFT correctly describes the overall nature and character of the electronic
bands in pnictides. It is worth noting that the evidence so far is still
incomplete and there remain open questions as regards the detailed
comparison of, for instance, the calculated anisotropy and exact shape of
the M-pocket in some compounds (see, e. g., V. B. Zabolotnyy et al,
arXiv:0904.4337). These details, however, remain beyond the scope of our
semiquantitative discussion.

\bibitem{note} There has been evidence that the electron states in Ba(Fe,Co)%
$_{2}$As$_{2}$ are subject to larger transport relaxation rates than the
hole states\cite{Wen,Paris}. This only strengthens our case, since we want
the bulk current to be dominated by electrons.

\bibitem{Wen} L. Fang, H. Luo, P. Cheng, Z. Wang, Y. Jia, G. Mu, B. Shen, I.
I. Mazin, L. Shan, C. Ren, and H. H. Wen, Phys. Rev. B \textbf{80},
140508(R) (2009).

\bibitem{Paris} F. Rullier-Albenque, D. Colson, A. Forget, and H. Alloul,
Phys. Rev. Lett. \textbf{103}, 057001 (2009).

\bibitem{note2} For a thin barrier a more relevant quantity than   $%
\left\langle N(E_{F})v_{F}^{2}\right\rangle $ is $\left\langle
N(E_{F})v_{x,y,z}\right\rangle ,$ but in case of FeSe it has qualitatively
the same anisotropy and therefore is not shown in Fig. \ref{vel}.
\end{thebibliography}
\end{document}